# Engineering Interfaces with Electric and Magnetic Fields: A Comprehensive Review.

[1*]Priyanka Biswas,[2]Gaurav Pransu

[1]Department of Science, Alliance University, Bengaluru 562106, India
[2] Institute of Physics, Bijenička cesta 46, 10000 Zagreb, Croatia

* priyanka.biswas@alliance.edu.in

The spontaneous organization of particles at the boundary between two fluids is a common occurrence. Scientists have extensively investigated various internal and external factors to manipulate and guide these self-assembly processes. This review article centers on the impact of external fields, particularly electric and magnetic fields, on the self-assembly of intermediate-sized particles at interfaces. By modifying the shape and dimensions of particles, attractive forces can be precisely controlled. Furthermore, the intensity and orientation of external fields exert a substantial influence on the resulting assembled structure. These systems hold immense promise for groundbreaking applications in drug delivery, catalysis, and other transport mechanisms.

## 1 Introduction

Over the years, self-assembly has become a prominent term in soft matter science. It refers to the spontaneous formation of ordered structures from units such as molecules, colloids, or micelles, without direct external influence. This phenomenon is of great interest because it is energy-efficient and can occur in multiple steps, enabling a convenient 'bottom-up' approach for creating novel functional materials [1]. The concept is deeply rooted in the evolution of life from unicellular organisms. Advances in technology have allowed this process to be replicated in various fields, such as electronics [2], photonics [3], and memory devices [4], to build hierarchical structures from micro- or nanoscale units. In colloidal chemistry, substantial literature has emerged over recent decades on colloidal self-assembly [5]. Understanding the intrinsic factors that govern particle self-assembly is essential for achieving the desired structure and functionality of nanoscale devices. These factors include interparticle interactions like electrostatic and van der Waals forces, as well as extrinsic factors such as external fields and capillarity [6]. Particle shape, size and surface charge are some of the factors influencing the interactions. The energetics of these interactions are comparable to the thermal energy of the surrounding medium, enabling the system to autonomously explore the energy landscape until it reaches a stable configuration.The introduction of external fields changes this dynamic. Field-assisted assembly techniques have emerged to provide precise control over the assembly process, resulting in desired structural hierarchies. Despite this, self-assembled structures under specific conditions tend to be low-energy configurations. Trapping nanoparticles at fluid-fluid interfaces is a time-tested and promising technique for inducing order in colloidal systems. This technique lowers interfacial energy, causing particles to become irreversibly trapped. This behavior is particularly valuable for stabilizing emulsions and foams. Historically, Ramsden [7] and Pickering [8] demonstrated how micrometric colloidal particles could stabilize paraffin-water emulsions by reducing interfacial tension, leading to 2D colloidal crystal formations. These emulsions were subsequently named after Pickering. Later, Pieranski [9] investigated the patterning of polystyrene particles at air-water interface into 2D triangular lattice structures due to dipole interactions. The energy required to desorb a particle from the interface varies with its size, shape, and interactions with other particles. Notably, nanoparticles, due to their lower desorption energies relative to thermal energy, are more easily displaced, resulting in particle exchange at the interface.

One of the ways to lock-in these particles into desired arrangements is by external fields if they are receptive to them, such as magnetic nanoparticles by an external magnetic field, charged particles by an electric field, thermal gradient, etc.[10] The entrapment and symmetry-breaking at interface play a crucial role in regulating the response of the particles to the external field. The responsiveness of the particles to the external field is again significantly dependent on surface charges, shape, and sizes. This review is concerned with the status of research on field-assisted self-assembly of particles at fluid interfaces. The article gives both a unified overview of recent experimental advances and theoretical studies in this regard, starting with discussing the cases for single particles (both spherical and non-spherical shapes) followed by interparticle interactions at interfaces.

## 2 ISOLATED PARTICLES AT FLUID-FLUID INTERFACE

To understand how trapped particles behave at the interface, we begin by examining an isolated particle. Analyzing the impact of a single particle's adsorption on the interface helps us generalize to multiple particles and evaluate the effect of external fields. Consider a particle suspended near a planar fluid-fluid interface. When this particle attaches, it removes a patch of fluid-fluid contact, $\Delta A_{P1}$ corresponding to the area displaced by the particle from fluid 1, making a gap in the interface of area $\Delta A_{12}$.[21] Additionally, the particle can induce distortions in the surrounding interface with an area $\delta A$. The net energy change or trapping energy $\Delta E$ at the interface is given by[38]

$$\Delta E = (\gamma_{P1} - \gamma_{P2}).A_{P1} + \gamma_{12}\Delta A_{12} + \gamma_{12}\delta A \quad (1)$$

where $\gamma_{P1}$ and $\gamma_{P2}$ are the surface energies per unit area of the particle-fluid 1 and particle-fluid 2 interfaces respectively and $\gamma_{12}$ is the surface tension between fluid 1 and 2. The difference in free energy between the two states (before and after particle trapping) dictates the preferred orientation for physically anisotropic particles, which may have multiple possible orientations[22]. This principle can be used to predict the orientation of ellipsoids, dumbbells [11], and other anisotropic shapes. Well-known anisotropic particles include disks [12], cubes [13], and Janus or patchy particles, which have regions with different chemical or physical properties [14]. When the trapping energy is significantly larger than $k_BT$ the particle remains trapped and cannot spontaneously detach from the interface. Trapping particles at fluid-fluid interfaces is a highly effective method for nanoparticle assembly, driven primarily by the reduction in interfacial energy, with lateral capillary forces further controlling the assembly process. In this context, we will explore spherical particles, which represent the simplest symmetry. As we proceed, we will also address the increasing complexity of the energy expression for particles with reduced symmetry.

## 2.1 *Spherical particle*

As mentioned earlier, Pieranski [9] in 1980 drew attention to the system of spherical PS particles forming crystal-like ordering at flat interfaces. To have a clearer picture of the underlying forces driving the assembly, we focus on one spherical particle of radius $r$ resting at a flat interface between fluid 1 and fluid 2. Let $\theta_c$ be the contact angle (Fig. 1). The particle can attach without deforming the surrounding interface, making $\delta A = 0$, with the contact line forming a circle in the plane of the interface. Disregarding any other external forces and considering a finite interface thickness, the net trapping energy is:[38]

$$\Delta E = -\gamma_{12}\pi r^2(1 - |\cos\theta_c|)^2 \quad (2)$$

Where $\cos(\theta_c) = {(\gamma_{P1} - \gamma_{P2})}/{\gamma_{12}}$.

When a particle attaches to the fluid-fluid interface, it reduces the interfacial area and lowers the system's energy, influenced by the particle's wetting properties. Perfectly smooth spheres with uniform wetting conditions leave the surrounding interface undisturbed, resulting in high trapping energy. In contrast, partial wetting causes interface distortion. For example, at an air-water interface with surface tension $\gamma_{12} = 72 mN/m$ or $18 k_BT/nm^2$ typical trapping energies for microparticles can be $10^5 - 10^6 k_BT$ compared to thermal energy (a few $k_BT$), leading to strong trapping. Nanoparticles, however, have lower trapping energies of $10 - 10^2 k_BT$ making them more prone to thermal displacement from interfaces. Contact-line pinning, due to nanoscopic roughness or chemical heterogeneity, also affects trapping energy by distorting the interface, leading to $\delta A \neq 0$ in Eq. 1 [15] For isolated particles, pinning changes the trapping energy by $\gamma_{12}\delta A$; termed the self-energy contribution [16]. Evaluating the trapping energy for spherical particles with pinned contact lines is complex due to unknown factors, such as the angle characterizing the particle's immersion in the fluid, which replaces the contact angle in Eq. 2. The self-energy term $\delta A$ in Eq. 1 represents the area of particle-induced deformation in the interface. $\delta A$ can be calculated by decomposing the contact line shape into Fourier modes, considering the interface height around the particle above the reference plane, and evaluating the area for each mode. Similar terms arise from higher-order modes in the multipole expansion.[38]

## 2.2 *Non-spherical particle*

The majority of interface assembly studies involve spherical particles. However, with significant advancements in synthesizing anisotropic nanoparticles, these particles have become unique subjects of study at interfaces. Anisotropic particles have been shown to enhance light control in photonic crystals [17] and serve as efficient nanocarriers for drug delivery [18]. They also exhibit intriguing behavior when dispersed in nematic liquid crystals [19] and demonstrate non-Newtonian rheological properties [20]. Unlike spherical particles, non-spherical particles must distort the interface to maintain a constant contact angle, invalidating the assumption of a planar fluid interface. The contributions to trapping energy for complex-shaped particles remain consistent and include the reduction in interfacial energy due to the particle's presence ($\Delta A_{12}$), modified by the particle's wetting properties, and the energy associated with interface distortion ($\delta A$). Accurately determining each contribution is challenging due to the lack of analytical solutions for the equilibrium wetting configuration of complex-shaped particles [21, 22]. Consequently, factors such as the size of the hole in the interface, the particle height, and the extent of interface distortion are usually obtained through simulations. For visual reference, these aspects are illustrated in Fig. 1. As with spherical particles, the contact lines around non-spherical particles can also be decomposed into Fourier modes.[38] For elongated particles, expansion is analyzed in ellipsoidal coordinates based on aspect ratio, with quadrupolar modes describing the far-field interface shape. A phase diagram can be created to show configurations of isolated cylinders. The free energy dependence on line tension τ is given by Eq. 3. Bresme et al. [23] found that, without an external field, particles favor a parallel orientation to maximize area away from the fluid-fluid interface. The stability difference between parallel and perpendicular orientations increases with aspect ratio, making perpendicular configurations less stable as aspect ratio grows. For a nanometric particle with a 2:1 aspect ratio, trapping energies are approximately$10^2$ and $2 \times 10^2 k_B T$ for perpendicular and parallel configurations, respectively. At an aspect ratio of 10:1, the energy difference is about$10^3 k_B T$ making perpendicular orientation unlikely. Low surface tension reduces the free energy difference, suggesting that external fields could stabilize vertical orientations. Anisotropic particles attach to interfaces similarly to spheres. Wang et al. [24] showed that ellipsoidal microparticles roll into the interface due to contact line displacement. Complex particles adopt various configurations, with isolated particles orienting to maximize interface disruption. Rod-shaped colloids are popular for their functional properties and exhibit phase behavior in bulk and orientation at interfaces based on factors like aspect ratio and wettability. The general trapping energy expression for ellipsoidal or rod-like particles at a fluid-fluid interface is given by [27].

$$\Delta E = (\gamma_{P1} - \gamma_{P2}).A_{P1} - \gamma_{12}\Delta A_{12} + \tau L \quad (3)$$

where $\tau$ is the line tension of the three-phase line, and $L$ is the length of the three-phase line. The relevant parameters are depicted in Fig. 1. Line tension becomes an important factor for small micrometer-sized and large nanometer-sized particles. Line tension may qualitatively alter the behaviour of non-spherical particles at interfaces causing them to undergo orientational changes to minimize the length of the contact line. Faraudo [27] derived expressions for the free energies of ellipsoids in two orientations: with the major axis parallel to the interface plane l ($\Delta E_{parallel}$) and with the major axis perpendicular to the interface ($E_{perp}$), as follows [23]

$$\Delta E_{parallel} = 2\pi r L[\pi(\gamma_{P1} - \gamma_{P2}) + \theta_c(\gamma_{P2} - \gamma_{P1}) + \gamma_{12}\sin\theta_c] \quad (4)$$

$$\Delta E_{perp} = (\gamma_{P2} - \gamma_{12} - \gamma_{P1})\pi r^2 + (\gamma_{P2} - \gamma_{P1})2\pi r h \quad (5)$$

where $r$ is the cross-sectional radius of the ellipse and $h$ is the penetration depth of the nanorods in the water when assembled perpendicular to the interface (Fig. 1). Thus, isolated nanorods orient parallel to the plane of the interface to maximize the interfacial coverage per particle and minimize the Helmholtz free energy. The interfacial distortions around ellipsoids and cylinders have quadrupolar geometry.[28, 21, 22, 38] As a special case, of multipolar symmetry, cuboidal particles show that, in certain orientations, particles excite hexapolar, or octupolar modes, aside from quadrupoles in the interface.[29]

# 3 INTERPARTICLE INTERACTION AT THE INTERFACE

The conditions and constraints of particle assembly are determined by the balance between attractive and repulsive forces, similar to self-assembly in bulk colloids. Common forces involved in self-assembly include attractive van der Waals forces, electrostatic double-layer repulsion, hydration repulsion, depletion forces, and hydrophobic interactions [6]. A system is considered kinetically stable when these opposing forces are comparable in magnitude and vary with particle separation. At interfaces, when conditions favor particle adsorption, the interactions among particles become more confined to the interface. Particle interactions become more favorable along the interface but less so through the phases on either side [30]. Self-assembly at fluid interfaces is influenced by both direct interactions, such as van der Waals and electrostatic forces, and indirect interactions, including forces from interface distortions or capillary effects [23, 31]. A liquid bridge between two solid surfaces generates capillary forces, which can be attractive or repulsive depending on the curvature of the liquid interface. As particles approach, their interface distortions interact, leading to the formation of meniscus regions that overlap and reduce interface slope and distortion area $\delta A$ These interactions, known as lateral capillary forces, become stronger with increased interfacial deformation. Similar particles on a liquid interface tend to attract each other due to gravity-induced meniscus deformation, especially for small anisotropic particles affected by electrostatic stresses [32]. The flotation capillary force arises from the particle's weight and the Archimedes force, and for nanometric non-spherical particles, this force can be significant due to meniscus deformation caused by electrostatic stresses [33, 90]. Capillary attraction forces emerge when particles are partially submerged and restricted within a liquid layer, as opposed to freely floating. [34, 35,36] In this situation, the deformation of the liquid surface is determined by the wetting characteristics of the particle surface. [40] This force is known as immersion capillary force. These two kinds of forces can be attractive or repulsive depending on the meniscus slope of the contact lines of the two particles. Capillary interaction energies between two particles can be calculated based on the difference between the changes of the fluid-fluid interface area for placing two particles at a given center-of-mass separation and for placing two isolated particles at a planar interface. The corresponding capillary interaction energy is [31],

$$\Delta E_{cap} = \gamma_{12}(\Delta A_{AB} - \Delta A_A - \Delta A_B) \quad (6)$$

where $\Delta A_{AB}$ is the change of the interface around the interacting particles, $\Delta A_A$ and $\Delta A_B$ are the changes of the interface areas around the single particles. Interface-mediated interactions follow power laws in the far-field and are thus long-ranged. In the near field, the interactions strongly depend on the particle shapes. For a system of $N$ randomly oriented colloidal particles dispersed at a fluid-fluid interface, the (free) energy at the interface is given by [29, 37],

$$E_N(\Omega) = \gamma_{12}[S(\Omega) - A_{12} + A_{P1}(\Omega)\cos\theta_c] \quad (7)$$

where $\Omega$ represents the configuration of the particles. In general, any arbitrary anisotropic shape, the position and orientation are addressed by six parameters in 3D (cartesian positional and angular) so $\Omega$ is a function of 6N entries. When there are no particles adsorbed at the interface $S(\Omega)$ is simply the total interfacial area $A_{12}$.[75] However, when particles are trapped, the total fluid-fluid surface area alters (analogous to $\Delta A_{AB}$ for a two-particle system in Eq. 6). The total solid surface area of the particles wet by fluid 1 is $A_{P1}(\Omega)$ with a contact angle $\theta_c$, $E_N$ can be computed numerically for any configuration $\Omega$ in the near field, however analytical solutions for the interaction forces are available in the far field.[ 31, 37]

The strong deformation fields around anisotropic particles make them excellent vehicles for studying capillary interactions. The dynamics of microparticle assembly were first observed for ellipsoids at a water–oil interface a decade ago,[28] motivating interest in ellipsoidal particle assembly.[22, 39] The particles, with a major axis $10\mu m$ and minor axes $2\mu m$, interacted over distances as great as six particle lengths with weakly Brownian trajectories in the far-field, and well-determined paths in the near field. Particles approached in either tip-to-tip and side-to-side configurations. The former requires more energy than the latter. Tip-to-tip interactions exhibit quadrupolar capillarity in the far field, while side-to-side arrangements show significant contributions from higher-order modes at greater distances, according to the simulations. Simulation of weakly nonspherical particles assuming pairwise additivity suggests that even nanometric deviation from sphericity can drive capillary assembly of ellipsoids into a variety of structures that include dendritic-trapped con- figurations, rafts, and hexagonal lattices.[23, 38] Numerical calculations for two-particle, three-particle, and four-particle interactions predict a variety of stable and metastable configurations.[41]. Orthorhombic particles are a classic example where multiple stable orientations were found, depending on the particle aspect ratio and contact angles.[42] Anisotropic particles align as they transition to preferred configurations, with interactions initially driven by their quadrupolar modes. As they approach each other, higher-order modes, near-field distortions, and features like corners and edges influence the interaction strength and closest approach distance. Additionally, adjustments in wetting configurations and the associated solid–liquid wetting energies also contribute.[38]. He *et al.* demonstrated that by controlling the interfacial energy between different liquids and suspended CdSe nanorods, the aspect ratio of the nanorods, and their concentration, the lateral packing of the nanorods can be varied. [43] The proximity or packing of the particles is limited by particle roughness. Wavy contact lines pinned on the rough sites would create local disturbances near the particle, which is important only in the very near field.[44] When neighboring particles approach, these disturbances would interact. If they matched perfectly, with identical wavelengths, phases, and amplitudes, particles would attract. However, if they differed, as would be expected for random roughness, particles would be repelled. This concept was recently demonstrated using particles with wavy edges.[45] In the far field, these particles experience the usual capillary attraction. However, when distortions from the wavy contact lines overlap, particles with differing undulations are repelled. All of these have been discussed in prior literature.[46, 31, 37, 22, 38] As a bottom line, it can be said that depending on the particle orientation, position, and shape of the interface, subject to appropriate boundary conditions at the three-phase contact line, the fraction of the particles immersed in each phase can be established. These values have a significant effect on bulk fluidic properties such as viscous drag, radiation pressure force in optical trapping experiments, etc.[22,47]

# 4 EFFECT OF EXTERNAL FIELDS

The use of external fields to control particle suspensions has long been a powerful means for tailoring the mechanical, optical, and electronic properties of materials in various research[48] and applications.[49] They have also been used as efficient routes for self-assembly at interfaces.[25, 51, 52] The reason for this is that in a stable colloidal system, the particles are charged to hinder agglomeration, and the adsorption of particles at the interface is dynamic. This means that there is an inter-particle repulsion along with repulsion between the particle and fluid-fluid interface. As a charged particle comes close to a fluid-fluid interface, it faces repulsive image-charge forces, causing significant distortion of the particle's electrical double layers due to the interface's proximity

[53] If such particles can be electrically or magnetically actuated, particle-interface contact becomes easier. More recently, external fields have emerged as key methods to direct the assembly of colloidal and nanoparticles. Electric or magnetic fields are obvious candidates for field-directed assembly. Colloidal and nanoparticle assembly in fields occurs due to induced interactions. Field-induced interactions in suspensions have been studied in detail for both magnetically and electrically polarizable colloids, due to their importance to magneto- and electrorheology.[50]

Piao et al. [54] investigate the development of stimuli-responsive polymer/inorganic hybrid materials via Pickering emulsion polymerization and analyze the rheological properties of smart fluids under electric (electrorheological) and magnetic (magnetorheological) fields. Field-responsive particles interact with an external field due to their inherent permanent dipole, aligned with the particle's major axis. When a field is applied perpendicular to the interface, it generates a torque that aligns the particle with the field, causing rod-like or ellipsoidal particles to shift from a parallel to a perpendicular orientation. This external field contributes to the particle's free energy at the interface. Assuming the external field is expressed as $F_{ext} = -\bar{m}.\bar{H}$ where $\bar{m}$ is the dipole moment and $\bar{H}$ is the field strength respectively. If $\varphi$ is the angle between the field and dipole vectors, the free energy of the particle at the interface can be written as [23]

$$\Delta E = (\gamma_{P1} - \gamma_{P2}).A_{P1} - \gamma_{12}\Delta A_{12} + \tau L - mH\cos\varphi \qquad (8)$$

$\varphi = 0$ indicates the particle is in the perpendicular (or vertical) orientation, while $\varphi = \pi/2$ indicates it is in the parallel (or horizontal) configuration. Line tension significantly affects the wetting properties of nanoparticles mainly when interfacial tension is very low and the three-phase line has high curvature [55]. The term related to the displaced interfacial area ($A_{12}$) accounts for the eccentricity of ellipsoids, making the energy dependent on particle orientation. Less anisotropic particles (lower aspect ratio) show a greater field effect. For particles with low anisotropy, the perpendicular orientation aligns with the field. In the case of spherical particles, the vertical orientation is preferred even with minimal external field strength. The external field thus imposes an orientation preference on the particle. For particles with intermediate anisotropy, the field causes the particle to tilt at an angle θ to the interface.[23] The transition from horizontal to vertical orientation occurs abruptly at a critical field strength, which depends on the particle's size, aspect ratio, and interfacial tension [26]. Fig. 2 illustrates the relevant terms for spherical and cylindrical particles at fluid interfaces.

## 4.1 *Magnetic field-induced assembly*

Magnetic particles in an external field experience two types of forces; namely, the dipole-dipole interactions between particles and the dipole–field interaction. [56] The resultant potential between the magnetic moments $\overline{m_i}$ and $\overline{m_j}$ is given by[58],

$$U_{dd}(\bar{d}, \overline{m_i}, \overline{m_j}) = \frac{\overline{m_i}.\overline{m_j}}{d^3} - 3.\frac{(\bar{d}.\overline{m_i})(\bar{d}.\overline{m_j})}{d^5} \qquad (9)$$

$\bar{d}$ denotes the relative position of the dipoles. The intensity of the dipole-dipole magnetic interaction between two identical particles, compared to thermal energy, is quantified by the magnetic coupling parameter $\Gamma_m$ which is defined for two identical magnetic moments as [101]

$$\Gamma_m = \frac{\mu_0 m^2}{16\pi r^3 k_B T}$$

where $\mu_0 = 4\pi \times 10^{-7} NA^{-2}$ is the free space permeability and $k_B T$ is the thermal energy, where $k_B$ is Boltzmann's constant and $T$ is absolute temperature. For $\Gamma > 1$, the dipole-dipole interaction induces the formation of aggregates, due to precedence of magnetic coupling over thermal randomization. For spherical particles with residual magnetization, the dipole-dipole potential causes the formation of linear, ring-like, and cluster-like structures even in the absence of an external magnetic field and at moderate temperatures.[57] When a uniaxial magnetic field is applied to these systems, it generates a torque that aligns and reorganizes the self-assemblies in the direction of the field. Superparamagnetic particles, however, only display magnetic dipole-dipole interactions when an external field is present. If the induced dipole moments are on the same plane as the relative position vector $\bar{d}$ then $U_{dd}(\bar{d}) = \frac{\mu_0 m^2}{4\pi d^3}(1 - 3\cos^2\alpha)$
where $\alpha$ is the angle between the dipole and the line connecting the dipoles which can range from 0 – 90°.[60] The corresponding force is given by $F = \frac{3\mu_0 m^2(1-3\cos^2\alpha)}{4\pi d^4}$. The dipole–dipole interaction is attractive when $\alpha$ is below 54.098° and repulsive when $\alpha$ varies from 54.098° to 90°.When two dipoles are aligned head-to-end, the dipole interaction is attractive, $F_{att} = -\frac{6m^2}{d^4}$. While they are aligned side by side, the net interaction is repulsive given by $F_{rep} = \frac{3m^2}{d^4}$. [58] Fig. 3 shows the various forces between the magnetic particles.

At the critical angle of 54.098° the interaction approaches zero and Brownian motion tends to randomize the colloidal particles. However, when the dipole interaction energy is at least one order of magnitude greater than thermal energy the particles self-assemble into chain-like structures under the influence of external magnetic field. Along the magnetic field, the particles attract ($F_{att}$) each other and form chains due to head-to-end alignment of dipoles while the repulsive interaction ($F_{rep}$), results from the side-by-side configuration of dipoles, keeping the chains apart.[58] Thus tailoring of the spatial magnetic field allow control over both the global and local assembly behavior.[59] When fields rotate in the plane, the average dipolar interaction energy becomes attractive if the field rotates at a high frequency, with the rotation periods being much shorter than the typical rotational and translational times of the adsorbed particles. If the field precesses at a high frequency around an axis perpendicular to the interface, with a precession angle $\alpha_p$, the dipole-dipole potential within the interface plane is given by,[101]

$$\langle U_{dd} \rangle = -\frac{\mu_0 m^2 (1 - 3\cos^2 \alpha_p)}{4\pi d^3}$$

#### **4.1.1** *Spherical particles*

Magnetically responsive particles adsorbed at interfaces can be controlled by an external magnetic field when the magnetic energy is on par with or exceeds the surface energy. However, the adsorption at the interface does not notably affect the long-range magnetic interactions between particles, as these interactions are not screened or modified by changes in particle distribution. The external magnetic field generates a magnetic torque $\overline{\tau_m} = \overline{m} \times \overline{H}$ which is balanced by a viscous torque resulting from contact line friction. The interfacial torque, given by $\tau_{int} \propto \gamma_{12} r^2 \alpha_t$ where $\alpha_t$ is the torsion angle, [60] describes the force exerted at the interface. When the magnetic energy surpasses the surface energy—typically under conditions of low fluid-fluid interfacial tension, high magnetic field strength—the adsorbed particles rotate about an axis parallel to the fluid interface. Since the interfacial torque scales with the square of the particle size, smaller nanoparticles are more readily reoriented at the fluid interface. Nonetheless, smaller particles often have weaker permanent magnetic moments and a reduced response to magnetic fields. This limitation can be overcome with sufficiently strong magnetic fields.[61] For ferro/ferrimagnetic particles at fluid interfaces, the interfacial torques usually outweigh the magnetic torques. Orienting the magnetic field perpendicular to the interface helps to prevent out-of-plane rotation of the particles[46,101, 73]

An oscillating out-of-plane field (having frequency $\omega$), $He^{i\omega t}$ induces rotational motion in the particles, producing a magnetic torque $\tau_0 e^{i\omega t}$ and causing deformation of the fluid interface. This displacement of the three-phase contact line due to the oscillatory rotation is represented by $\varphi_0 e^{(i\omega t + \delta\omega)}$. As a result, the adsorbed particles rotate in sync with the applied field.[61]. A novel method of trapping magnetic particles by magnetic tweezers to probe the wetting at the interface as a result of the torques generated was proposed by Cappelli *et al.*[61] The angular orientation of the particles was studied as a function of time when the particles are subjected to varying torques[102]. At lower torques, the particle deforms the interface up to the point where it overcomes the contact line pinning. Such kind of wetting phenomena is known as dynamic wetting, as opposed to static wetting for the trivial case. This limiting configuration depends on the particle surface properties and interfacial tension. At higher torques, the contact line displaces at different rotational speeds. Cappelli[62] demonstrated an interfacial rheometry technique based on periodic attraction and repulsion between pairs of micrometer-sized magnetic particles regulated by controlling the dipole forces by Intra-Pair Magnetophoresis (IPM). [63, 103] The interfacial drag coefficient, $\zeta_r$ was obtained from the particle displacements ( which is in $\mu m$), forces ($pN$) and velocities ($\mu m/s$) using the methodology mentioned in van Reenen *et al.*[63] When the magnetic field is applied normally to the interface, the force between particles at close proximity is repulsive.[102] In the low Reynolds number regime (Re << 1), hydrodynamic drag limits the dipole-dipole interaction. In this limit, the magnetic torque is counteracted by the hydrodynamic torque, described by $mH \sin\varphi = \zeta_r \varphi$. Solving this equation reveals the time-dependent relationship between the angle of the magnetic moment and the applied field When the field is changed to in-plane configuration (parallel to the interface) then dipole-dipole interaction becomes attractive. The application of stronger magnetic fields or higher magnetic susceptibility of the particles revealed non-linear responses of the interface.[62] Using the competitive balance of the magnetic and interfacial forces on the trapped spherical paramagnetic colloid, Tsai et al. reported a microfluidic interfacial tensiometry technique to measure ultralow interfacial tensions up to $10^{-5} - 10^{-6} Nm^{-1}$).[64] The assembly at the interface in such a case is reversible, as reported by Martinez et al. for linear aggregates of micron-sized magnetic particles. Disassembly of the particles occurs when the applied field is abruptly tilted out of the confining surface (out-of-plane motion). The chains of laterally aggregated particles "unzips" causing partial fragmentation of the chains and gradual separation of the monomers and the abrupt colloidal explosion. The assembled structures disintegrate due to thermal fluctuations as soon as the applied field is switched off.[52] 2D ordering of supermagnetic colloidal particles at the interface shows two-fold melting transitions were observed from solid to liquid phase with an intermediate hexatic phase as described by Zahn et al. [51]. Similarly, Wen et al.[65] reported 2D colloidal crystal ordering by applying perpendicular fields. The balance between the repulsive magnetic interaction and the attractive interaction due to the weight of the particles projected along the surface tangent, yielded triangular, oblique, rectangular and square lattice structures. By using two different-sized magnetic particles, local formations five-fold symmetrical structures were also achieved. These lattice structures were tunable by adjusting the polar and azimuthal angles of the magnetic field relative to the surface normal. Grosjean et al. [66] explored triangular lattice formations for millimeter-sized particles on a fluid interface, where capillary attraction on a curved interface is balanced by dipolar repulsion. Number of particles in the assembly as well as the amplitude and orientation of the applied magnetic fields play significant roles in deciding the formed structures.[66] When the field is directed perpendicular to the liquid-gas interface, the particles repel each other forming stable triangular crystals. By tilting the external field, the interaction becomes anisotropic, and a mutual attraction appears upon a threshold tilt angle. Various stable crystal lattices including rectangular, oblique, chain-like oblique, and rhombic structures were obtained by varying the tilt angle, the colloidal density, and the strength of the magnetic field. Lefebure reported the formation of chains and compact circular aggregates of magnetic nanoparticles under perpendicularly applied magnetic fields larger than 60 mT at air–water interface as a result of a balance between van der Waals and magnetic dipole–dipole interactions.[67] Maintaining a magnetic field gradient at liquid-air interface yields tunable hexagonally ordered clusters of macroscopic ferromagnetic particles as reported by Golosovsky et al.[68] The particles assemble into hexagonally ordered clusters with magnetically tuned lattice constants. Han *et al*.[69] reported self-assembly of synchronized ferromagnetic spinners at air-water interface under rotating magnetic field applied in-plane of the interface. When a low-frequency alternating field is applied parallel to the fluid interface, it induces the formation of pearl-like chains that either aggregate or fragment depending on the dimensionless Mason number, defined as:[101]

$$Mn = \frac{\eta \dot{\gamma}}{\mu_0 m^2}$$

which is the ratio of the viscous force and dipolar attraction and $\dot{\gamma}$ is the shear rate. [69] The interaction between the spinners yields dynamic phases and lattices which exhibit self-healing behaviour. The ordering and dynamics of the spinner lattice are tunable by varying the field characteristics. At high field frequencies, particles self-assemble into dynamic planar structures due to the isotropic potential described in Eq. 9[52]. Depending on the amplitude and frequency of the field, various aggregates form, including pulsating clusters, lines, and spinners. These spinners, influenced by the in-plane rotating field, self-assemble and rotate in unison. The resulting flow promotes the development of dynamic crystalline lattices with self-healing behavior .[69, 70] Notable work by Snezhko et al. demonstrated dynamic snake-like self-assemblies of ferromagnetic particles by application oscillating field normal to the air-water interface. The segments of the snake-like structure exhibit long-range antiferromagnetic ordering, whereas each of these segments is ferromagnetically interacting. They report an effective exchange interaction to explain this observation. [71].

#### **4.1.2** *Non-spherical particles*

Bresme and Newton et al. studied the influence of magnetic fields on cylindrical[72] and el- ellipsoidal magnetic particles[23, 60] at liquid interfaces. Bresme et al.[23] demonstrated a thermo-dynamical model which determines the orientation and free energy of ellipsoidal magnetic particles under magnetic field directed perpendicular to the interface. Aspect ratio, particle size and external field strength play crucial roles in this regard. A critical field strength exists at which a particle transitions abruptly from a tilted configuration, where the major axis (dipole moment) is at an angle to the interface, to an alignment with the external field. [84] Calculations indicate that this critical field strength is highly dependent on the nanoparticle's size, aspect ratio, and interfacial tension.[23] Monte Carlo simulations set up using orientation-dependent interaction between the nanoparticles and the liquid for varying magnetic fields, corroborated with the observation of discontinuous transition. Further improvements by considering the interface deformation as well was given by Newton et al.[60] and Davies[74, 75] using lattice Boltzmann method. Davies et al. [75] demonstrated that a strong magnetic field applied perpendicular to the fluid interface can effectively control the capillary assembly of micro ellipsoids. Their simulations indicated that the orientation of the ellipsoids is governed by the interface's surface tension, the intensity of the magnetic field, and the particles' magnetic moment. The dipole-dipole interactions can be neglected compared to the capillary interactions owing to the weak magnetization of the ellipsoids. Anisotropic structures termed as ”capillary caterpillars” were observed as a result of side-by-side packing of the particles due to the dynamically changing capillary strength by the external field.[84] The discontinuous switching behaviour of the particles is useful in applications of photonic systems that require dynamic control of optical properties such as in electronic readers[75, 76] For cylindrical particles, Newton[72] demonstrated a similar transition. However, in this case by tuning both the aspect ratio and contact angle of the cylinder, subjected to external magnetic field, multiple locally stable orientations were obtained through non-volatile transitions. The rotational motion of ferromagnetic nickel nanowires at the interface un- der magnetic field were used to probe the nonlinear micro-rheology of layers of protein lysozyme adsorbing at air-water interface.[77, 103] In that case, deviations from Eq. 9 will occur due to non-linear responses.[78] Unconventional shaped micro-rafts ($50\mu m$ radius) with sinusoidal edge-height profiles under rotating magnetic field applied parallel to the interface, generate interesting assembly configurations owing to the near-field repulsive and attractive capillary interactions competing with torque generated via the rotating field. This induces a programmable self-assembly depending on the rotational velocity and the peripheral profile of the magnetic rafts. [79]

## 4.2 *Electric field induced assembly*

Particles are polarized under an electric field, and the dipole-dipole interaction between them causes assembly, similar to the previous case of magnetic field. For a pair of identical spherical particles, the dipole-dipole interaction energy is

$$U_{dd}(\bar{d},\alpha) = m\beta E\left[\frac{3\,\mathrm{co}^{\;2}\alpha-1}{\left({}^{d}/_{2r}\right)^{3}}\right] = m\beta E\left({}^{2r}/_{d}\right)^{3} P_2(\cos\alpha) \quad (10)$$

If $\alpha$ is the angle as defined earlier, $\beta = \frac{\epsilon_p-\epsilon_s}{\epsilon_p+2\epsilon_s}$ where $\epsilon_p$and $\epsilon_s$ are the dielectric constant of the particle and solvent respectively, $m$ is the electric dipole moment of the particle in response to the electric field $E$. $P_2$ is the second Legendre polynomial. Electric coupling parameter $\Gamma_e$ can be defined as $\Gamma_e = \frac{\pi\epsilon_0\epsilon_s(\beta E)^2 r^3}{k_B T}$ where $\epsilon_0$is the free space permittivity. $U_{dd}$ is attractive if the line joining the two dipoles is parallel to the electric field whereas it is repulsive for perpendicular alignments. Solving the Laplace equation $\nabla^2 U_{dd} = 0$ the force is calculated to be $\frac{rF}{k_BT} = \frac{3}{4}\Gamma_e\bar{f}$ where $\bar{f} = \left(\frac{2r}{d}\right)^4\left[\widehat{e_d}\left(2f_{\|}\cos^2\alpha - 2f_{\perp}\sin^2\alpha\right) + \widehat{e_\alpha}\left(2f_{tor}\sin 2\alpha\right)\right]$where $f_{\|}$, $f_{\perp}$and $f_{tor}$ are the parallel, perpendicular and torsional force components respectively,[81] which are function of dielectric constants, particle size and radial distance. $\widehat{e_d}$ and $\widehat{e_\alpha}$ are the unit vectors for the radial and angular components. The interaction is attractive, $F_{att} = -\frac{3}{2}\Gamma_e\left({}^{2r}/_{d}\right)^4$ if the line joining the dipoles is parallel to the electric field whereas it is repulsive, $F_{rep} = -\frac{3}{4}\Gamma_e\left({}^{2r}/_{d}\right)^4$ for perpendicular alignments.[80] These terms are calculated considering point dipole limitation where the force components are unity.[81] If the particle and the solvent both polarize under electric field to the same extent (i.e. $\epsilon_p = \epsilon_s$) then $\beta = 0$ and consequently the interaction force goes to zero. when $\epsilon_p \neq \epsilon_s$, excess bound charge appears on the particle surfaces under the electric field, such that far from the particle it appears as a point dipole with dipole moment $m = 4\pi\epsilon_0\epsilon_s\beta E r^3$. [56,81] Other than that, mobile charges, such as those in an electrostatic double layer also respond to applied fields and can contribute to the polarization.[82] However in the far-field, the induced field surrounding a polarized particle takes the form of a dipole and leads to a strong dipole-dipole interaction between particles. When the interaction is sufficiently strong to overcome Brownian motion (the relative strength characterized by $\Gamma_e$), particles will initially start to assemble and consequently form large clusters with distinct ordered structuring. Ideally, the coarsening proceeds to form the lowest energy structure, a body-centered tetragonal (bct), hexagonally close-packed (hcp), or face centered cubic (fcc) lattice, depending on the particle concentration and field. [50, 83]

Particles trapped at interfaces experience an electrostatic force normal to the interface when subjected to an external electric field. As these particles become polarized, they also repel each other due to dipole-dipole interactions. The dielectric constants of the fluids above and below the interface create a mismatch, affecting the electric field distribution. In the absence of particles, the electric field is perpendicular to the interface, with a constant intensity in the upper and lower fluids but a discontinuous change at the interface. When particles are present, they interact through dipole-dipole forces. The balance between electrostatic interactions and capillary forces determines the assembly of particles at interfaces.For charged particles, Coulombic forces are also significant, and external electric fields can mobilize these particles efficiently, a phenomenon characterized by their electrophoretic mobility. In bulk, polarized particles tend to cluster in the direction of the electric field. However, at an interface, this electrostatic interaction becomes repulsive. The effect of electric fields on dispersed particles depends on both the properties of the particles and the surrounding medium. Electric fields also influence the contact angle through variations in voltage or electrowetting [84]. Changes in the contact angle cause particles to move perpendicular to the interface to meet the new angle requirements[99]. Parameters such as electric field strength, particle geometry (isotropic or anisotropic), and field frequency are crucial for controlling interfacial self-assembly. The mechanisms for polarizing particles differ between media and types of electric fields (direct or oscillating). Factors like the relative polarizability of the particles and solvent, as well as the distribution of mobile (e.g., dispersed ions) and immobile charges (e.g., covalently bound surface charges), significantly impact how particles respond to electric fields.[104]

#### 4.2.1 ***Spherical particles***

Aubry et al. demonstrated electrically controlled monolayer packing of spheres at interfaces when a DC electric field is applied perpendicular to the interface.[85] For an isolated particle at interface acted upon by an electric field, the force acting normal to the interface has been numerically computed to be

$$F_{ev} = r^2\epsilon_0\epsilon_U\left({\epsilon_L}/{\epsilon_U} - 1\right)E^2 f_v \qquad (11)$$

Where $f_v$ is a dimensionless vertical force coefficient, $\epsilon_U$ and $\epsilon_L$ are the dielectric constants of the upper and lower fluid in contact[99]. The factor in the bracket of Eq. (11) disappears in the case of an isolated particle submerged in a single fluid. Hence it does not experience a force field under bulk condition. $F_{ev}$ is also dependent on the particle radius and the dielectric constants of the fluids in contact. It may be directed along or opposite to the buoyant weight of the particles depending on these parameters. The lateral electric force acting on the particle due to its neighboring particle–particle interactions is[85]

$$F_{el} = \epsilon_0\epsilon_U\left({\epsilon_L}/{\epsilon_U} + 1\right)r^2E^2\left({r}/{d}\right)^4 f_l \qquad (12)$$

where $f_l$ is the dimensionless lateral force coefficient. $F_{eL}$ is repulsive. The repulsive force thus depends on the sixth power of particle radius *r* and on the fourth power of the inverse of the distance *d* between the particles. This force was shown to be stronger than the random Brownian force, indicating that an electric field can indeed be used to manipulate small particles within a two-fluid interface. The normal electric force displaces the particles at the interface, causing additional deformation. This deformation eventually results in an attractive lateral capillary force, which is expressed as: [30, 24, 25, 22,86 ]

$$F_{lc} = -\left(-\epsilon_0\epsilon_U\left({\epsilon_L}/{\epsilon_U} - 1\right)r^2E^2 f_v + {4}/{3}\,\pi r^3\rho_p g f_b\right)^2 \frac{1}{2\pi\gamma d} \qquad (13)$$

where $\rho_p$ is the particle density and $f_b$ represent the buoyancy force coefficient. The first part of the above expression is the vertical force $F_{ev}$ mentioned earlier.

Fig. 4. depicts the above-mentioned forces for a pair of spherical particles. The first term varies as the fourth power of the particle radius and the second term as its sixth power, making the contribution of the particle's buoyant weight for micron and nanometric particles negligible. Thus, it is unlikely for uncharged particles to self- assemble under mere capillarity. However, charged particles[87] or occurrence of irregular contact lines cause self-assembly at this dimension.[36] This restriction is overcome by an electric field of O($10^6$) V/m when the particles experience electric force normal to the interface of at least an order of magnitude higher than the random Brownian force, which is the order at which capillary forces become significant.[88] The repulsive $F_{eL}$ is short-ranged (varying as $d^{-4}$) whereas attractive $F_{Lc}$ is long-ranged(varying as $d^{-1}$).[99] Therefore, at the equilibrium distance for a fixed filed strength, the net force acting on the particle is zero. The force balance leads to the following dimensionless equilibrium separation $d_{eq}/2r$ between two isolated particles given by,

$$\frac{d_{eq}}{2r} = \frac{1}{2}\left(\frac{2\pi\epsilon_0\epsilon_U\left(\frac{\epsilon_L}{\epsilon_U}+1\right)\gamma E^2 f_l}{r\left(-\epsilon_0\epsilon_U\left(\frac{\epsilon_L}{\epsilon_U}-1\right)E^2 f_v+\frac{4}{3}\pi r\rho_p g f_b\right)^2}\right)^{1/3} \qquad (14)$$

Thus varying the electric field intensity, the equilibrium distances between particles vis-a-vis the assembly and monolayer packing at interface can be controlled. Aubry [85] reported an excellent correlation of the equilibrium distances between the experimental results of particles at oil-water interface along with theoretical calculations where the force coefficients are obtained numerically. [89] Two quantities, namely Bond and Weber numbers are used to define the force balance of the sphere at the interface given by $BO = \frac{gr^2\rho_L}{\gamma}$ and $WE = \epsilon_0\epsilon_U\left(\frac{rE^2}{\gamma}\right)$respectively. When $BO$ approaches zero, i.e. for small (nanometric) particle sizes, the interfacial deformation due to the particle weight is minimal. As radii approach $10\mu m$ the deformation become significant.[90] As mentioned earlier, in presence of a high enough electric field, $F_{Lc}$ for nanometric particles becomes significant even when the $BO$ approaches zero. [88, 24]. Liu et al. demonstrated that liquid water marbles can coalesce under the influence of a high DC electric field when stabilized by silica particles. The threshold voltage required for coalescence depends on both the stabilizing particles and the surface tension of the aqueous phase. The number of marbles coalescing increases proportionally with the applied voltage [91]. However, DC electric fields can induce strong currents in electrolytes, potentially degrading materials and disrupting the distribution of free charge carriers. Consequently, experiments with DC fields often use poorly conducting liquids or operate at low electric fields. To circumvent these issues, high-frequency AC electric fields are frequently employed to polarize only the solid dielectric particles, avoiding effects on the ionic double layer or capacitive charges since charge redistribution requires finite response time [90, 92, 104].The force experienced by a particle at an interface subjected to AC fields is analogous to that in DC fields. The time-averaged force in an AC field,[105] where EEE is the RMS value of the AC electric field, can be represented by Equation 11. When a spherical particle is exposed to a non-uniform electric field and its dielectric constant differs from that of the surrounding fluid, the resulting electric stress on the particle's surface generates a net force known as the dielectrophoretic (DEP) force, causing the particle to move. For particles small relative to the length scale over which the electric field varies, the field gradient is considered constant, and the time-averaged DEP force acting on a spherical particle in an AC electric field is given by [93]

$$F_{DEP} = 2\pi r^3 \epsilon_0 \epsilon_s \beta(\omega) E_{RMS} \nabla E_{RMS} \quad (15)$$

where $E_{RMS}$ is the RMS value of the field, $\beta(\omega)$ is the real part of the frequency dependent Clausius- Mossotti factor (similar to the $\beta$ defined earlier) given by $\beta(\omega) = Re\left(\frac{\epsilon_p^* - \epsilon_s^*}{\epsilon_p^* + 2\epsilon_s^*}\right)$ where $\epsilon_p^*$and $\epsilon_s^*$are the frequency-dependent complex permittivities of the particle and the solvent, respectively. The Weber number in this case becomes $WE = \epsilon_0 \epsilon_U \beta(\omega)\left(\frac{rE^2}{\gamma}\right)$. For a particle at a two-fluid interface, the DEP force differs from the expression on Eq. 15. The Clausius-Mossotti factor $\beta(\omega)$ depends on the dielectric constants of the particle and the two fluids involved, and on the orientation of the particle within the interface as well.

The third factor also depends on the particle's wettability and buoyant weight. Electric fields have been effectively used to manipulate the assembly of particles at the interfaces of Pickering droplets through the dielectrophoresis (DEP) effect. Nudurupati et al. [94] explored how a uniform AC electric field influences the distribution of particles on the surface of a drop immersed in a different immiscible liquid with a finite dielectric constant difference. When the drop's dielectric constant is higher than that of the surrounding liquid, particles accumulate at the poles of the drop, forming a ring-shaped region near the equator when the drop's dielectric constant is lower. This effect occurs because the electric field on the drop's surface is nonuniform, even when the applied field is uniform[94]. Their simulations confirm that the electric field strength peaks at the drop's equator when its dielectric constant is lower than that of the surrounding liquid, leading to particle accumulation there. Conversely, if the drop's dielectric constant is higher, the electric field strength is greatest at the poles, causing particle accumulation at the poles. When particles gather at the poles, the electric field strength required to destabilize the interface decreases due to reduced interfacial tension. [94] Hwang et al. demonstrated that, in addition to redistributing particles on drop surfaces, electric fields can also destabilize Pickering emulsions. Their work showed that nearby droplets could merge through gaps on the drop surface left uncovered by particles [95]. In subsequent research, Nudurupati et al. [96] examined how the dielectrophoretic force (FDEP) depends on the radii of particles and drops, as well as the dielectric properties of the fluids and particles. When the drop radius exceeds a critical size, particles cannot be concentrated on the interface, as the drop breaks at a lower electric field intensity than required for particle concentration. FDEP inversely correlates with drop radius, making spatial concentration manipulation more effective with smaller drop sizes. Particles undergoing positive dielectrophoresis can be segregated from those experiencing negative dielectrophoresis, leading to a distribution where one type accumulates at the poles and the other at the equator. Additionally, the electric field intensity can be tuned to eliminate specific types of particles from the drop surface. [96] Dommersnes et al. [97] demonstrated that electric fields induce the formation of colloidal "armour" on oil drops, organizing into 'equatorial' ribbons and 'longitudinal' dipolar chains. This phenomenon is attributed to electrohydrodynamic circulation within the drop and dipole interactions between particles [98]. The width of the armour can be actively controlled by adjusting the electric field strength, which is closely related to the polarizability contrast between the particles and the drop oil, as well as the conductivity of the colloidal particles[97].

#### 4.2.2 *Non-spherical particles*

Crassous et al. [92] combined experimental and computational approaches to show that ellipsoidal particles can reversibly self-assemble into well-defined tubular structures when exposed to an AC electric field. Their model calculations demonstrated the influence of aspect ratio and field-induced dipolar interactions on particle assembly. When subjected to a sufficiently strong electric field, certain stable alignments of anisotropic colloids become unstable. For instance, rods that normally align with their long sides facing each other may shift to an unstable configuration where their ends touch, due to an electrostatic torque $\overline{\tau_e} = \overline{m} \times \overline{E}$ acting perpendicular to the interface. Janjua et al. [99] provided experimental evidence that an external electric field normal to a fluid-fluid interface effectively aligns rods floating on the interface and adjusts the lattice spacing of the formed monolayer.[105] In the presence of a strong electric field (5000 V), the rods in a monolayer align end-to-end, causing the lattice spacing of the self-assembled monolayer to increase which is independent of the rods' initial orientation.[99] Similar to spherical particles, rods on the interface experience an electrostatic force normal to the interface and become polarized under the electric field, interacting via dipole-dipole interactions. These interactions lead to repulsive forces and torques, with the equilibrium distance between rods determined by the balance of attractive capillary forces and repulsive dipole-dipole forces.

# 5 Conclusions and Outlook

This article provides a outline of research for field-assisted assembly of particles at fluid-fluid interfaces. Interest in interface-assisted assembly has grown manifold over the years. With the advancements in synthesis techniques for nanoparticles, investigations have been carried out for various shapes, sizes and surface functionalities. Particle shape plays a significant role in deciding the energy landscape at the interface. We first discussed the case of isolated particles at interface. Deviation from sphericity induces various stable and metastable states of the particles at interface. The most reported systems are those of rod-like or ellipsoidal particles. Interparticle interaction between the particles is again shape dependent. Interfacial deformation caused by anisotropic particles are multipolar in nature owing to the orientation and position of the particles. The overlap of the deformations of the interface in the near field modifies the three-phase contact line that results in capillary interactions. For shape anisotropic particles, higher order deformations such as hexapolar or octupolar are obtained for certain particle orientations. The important physical aspects of the above-mentioned categories have been highlighted along with reports and results from the relevant case studies. The presence of field acts as an additional tool for tuning the assembly at interface. In this regard, the two important external influences of electric and magnetic field have been discussed. First a general formalism of the effect of an external field is provided where the particle's inherent dipole moment interacts with the spherical and cylindrical particles. Then, the interaction of the particle with the external field has been demonstrated. The two cases of spherical and non-spherical particles are discussed along with relevant examples from literature. The use of external fields to assist assembly at fluid-fluid interfaces holds immense potential in terms of designing assembly at interfaces. Contrary to the traditional use of micrometric sized particles, nanoparticles have gained importance over the years for their ease of synthesis and functionalization. Novel synthesis methods allow physical properties such as microwave or optically responsivity to be incorporated into the nanoparticles. These particles then become markers with potential in investigation of collective behaviour at interfaces under various physico-chemical conditions. Moreover, biocompatible nanoparticles serve as biomarkers allow monitoring of drug assimilation at bio-interfaces or mem- branes. Magnetic nanoparticles have also shown to hold advantage for targeted drug delivery in a cargo pick-up-drop fashion [100] Nanoparticle shape is another interesting aspect that provide ample opportunities for further work. The short-range interactions could be modified by use of further anisotropic colloids and other hybrid constructs. Long-range electric and magnetic interactions can again be coupled in various degrees to further increase the complexity of the dynamical structures at the interface.

## Conflicts of Interest

There are no conflicts to declare.

### Acknowledgements

PB would like to thank Prof. Dillip Satapathy and Prof. Basavaraj M Gurappa for the discussions.

### Funding

This research did not receive any specific grant from funding agencies in the public, commercial, or not-for-profit sectors.

## REFERENCES

[1] George M. Whitesides and Mila Boncheva, Beyond molecules: self-assembly of mesoscopic and macroscopic components, PNAS 9 9 ( 2002) 4769-74. https://doi.org/10.1073/pnas.082065899

[2] John A. Rogers, Zhenan Bao, Kirk Baldwin, Ananth Dodabalapur, Brian Crone, V. R. Raju, Valerie Kuck, Howard Katz, Karl Amundson, Jay Ewing, and Paul Drzaic, Paper-like electronic displays: Large-area rubber-stamped plastic sheets of electronics and microencapsulated electrophoretic inks, PNAS 98(2001) 4835– 4840. https://doi.org/10.1073/pnas.091588098

[3] Chantal Paquet and Eugenia Kumacheva, Nanostructured polymers for photonics, Mater Today, 11(2008), 48-56 https://doi.org/10.1016/S1369-7021(08)70056-7

[4] W.A. Gifford and L.E. Scriven, On the attraction of floating particles, Chem. Eng. Sci. 26(1971), 287-297. https://doi.org/10.1016/0009-2509(71)83003-8

[5] R.H. Ottewill, A.B. Schofield, J.A. Waters and N. St. J. Williams. Preparation of core-shell polymer colloid particles by encapsulation. Colloid Polym Sci 275(1997), 274–283. https://doi.org/10.1007/s003960050081

[6] Jacob N. Israelachvili, Intermolecular and surface forces, 3rd edition, Academic Press London, 1991.

[7] W. Ramsden, Separation of solids in the surface-layers of solutions and 'suspensions' (observations on surface-membranes, bubbles, emulsions, and mechanical coagulation).—Preliminary account, Proc. R. Soc. London 1904. https://doi.org/10.1098/rspl.1903.0034

[8] Spencer Umfreville Pickering, CXCIV.—The interaction of metallic sulphates and caustic alkalis, J. Chem. Soc. Trans., 91(1907), 1981-1988. https://doi.org/10.1039/CT9079101981

[9] Pawel Pieranski, Two-Dimensional Interfacial Colloidal Crystals, Phys. Rev. Lett. 45 (1980) 569-572. https://doi.org/10.1103/PhysRevLett.45.569

[10] G. Sreenivasulu, V. M. Petrov, F. A. Chavez and G. Srinivasan, Superstructures of self-assembled multiferroic core-shell nanoparticles and studies on magneto-electric interactions, Appl. Phys. Lett. 105(2014) 072905. https://doi.org/10.1063/1.4893699

[11] Guido Avvisati, Teun Vissers, and Marjolein Dijkstra, Self-assembly of patchy colloidal dumbbells, *J. Chem. Phys.* 142(2015), 084905. https://doi.org/10.1063/1.4913369

[12] Junaid Qazi, Göran Karlsson, and Adrian Rennie, Self-Assembled Structures of Disc-Like Colloidal Particles, Prog. Colloid Polym. Sci. 138(2011) 61-65. https://doi.org/10.1007/978-3-642-19038-4_11

[13] Margaret Rosenberg, Frans Dekker, Joe G. Donaldson, Albert P. Philipse and Sofia S. Kantorovich, Self-assembly of charged colloidal cubes, Soft Matter, 16(2020), 4451-4461. https://doi.org/10.1039/C9SM02189B

[14] S. Jiang, Q. Chen, M. Tripathy, E. Luijten, K.S. Schweizer and S. Granick, Janus Particle Synthesis and Assembly, Adv. Mater., 22(2010), 1060-1071. https://doi.org/10.1002/adma.200904094.

[15] Anna Wang, Ryan McGorty, David M. Kaz and Vinothan N. Manoharan, Contact-line pinning controls how quickly colloidal particles equilibrate with liquid interfaces, Soft Matter 12(2016), 8958-8967. https://doi.org/10.1039/C6SM01690A

[16] Peter A. Kralchevsky and Kuniaki Nagayama, Capillary interactions between particles bound to interfaces, liquid films and biomembranes, Adv. Colloid Interface Sci. 85(2000), 145-192. https://doi.org/10.1016/S0001-8686(99)00016-0

[17] Krassimir P. Velikov, Alexander Moroz and Alfons van Blaaderen, Photonic crystals of core-shell colloidal particles, Appl. Phys. Lett. 80(2002), 49–51. https://doi.org/10.1063/1.1431698

[18] Z. Wang, Z. Wu, J. Liu and W. Zhang, Particle morphology: an important factor affecting drug delivery by nanocarriers into solid tumors. Expert Opin. Drug Delivery, 15(2017), 379–395. https://doi.org/10.1080/17425247.2018.1420051

[19] Maxim V. Gorkunov and Mikhail A. Osipov, Mean-field theory of a nematic liquid crystal doped with anisotropic nanoparticles, Soft Matter, 7(2011), 4348-4356. https://doi.org/10.1039/C0SM01398F

[20] Takehiro Yamamoto, Takanori Suga, and Noriyasu Mori, Brownian dynamics simulation of orientational behavior, flow-induced structure, and rheological properties of a suspension of oblate spheroid particles under simple shear, Phys. Rev. E 72(2005), 021509. https://doi.org/10.1103/PhysRevE.72.021509

[21] Eric P. Lewandowski, Marcello Cavallaro, Lorenzo Botto, Jorge C. Bernate, Valeria Garbin and Kathleen J. Stebe, Orientation and Self-Assembly of Cylindrical Particles by Anisotropic Capillary Interactions, Langmuir 26(2010), 15142-15154. https://doi.org/10.1021/la1012632

[22] Lorenzo Botto, Eric P. Lewandowski, Marcello Cavallaro, Jr and Kathleen J. Stebe, Capillary interactions between anisotropic particles, Soft Matter, 8(2012), 9957-9971. https://doi.org/10.1039/C2SM25929J

[23] F Bresme and M Oettel, Nanoparticles at fluid interfaces, J. Phys.: Condens. Matter 19(2007) 413101. https://dx.doi.org/10.1088/0953-8984/19/41/413101

[24] Anna Wang, W. Benjamin Rogers, and Vinothan N. Manoharan, Effects of Contact-Line Pinning on the Adsorption of Nonspherical Colloids at Liquid Interfaces, Phys. Rev. Lett. 119(2017) 108004. https://doi.org/10.1103/PhysRevLett.119.108004

[25] Eric P. Lewandowski, Jorge A. Bernate, Alice Tseng, Peter C. Searsona and Kathleeen J. Stebe, Oriented assembly of anisotropic particles by capillary interactions, Soft Matter, 5(2009), 886-890. https://doi.org/10.1039/B812257A

[26] Jorge L. C. Domingos, François M. Peeters, and W. P. Ferreira, Self-assembly and clustering of magnetic peapod-like rods with tunable directional interaction, PLOS ONE 13(2018) e0195552. https://doi.org/10.1371/journal.pone.0195552

[27] Jordi Faraudo and Fernando Bresme, Stability of particles adsorbed at liquid/fluid interfaces: Shape effects induced by line tension, *J. Chem. Phys.*, 118(2003) 6518–6528. https://doi.org/10.1063/1.1559728

[28] J. C. Loudet, A. M. Alsayed, J. Zhang, and A. G. Yodh, Capillary Interactions Between Anisotropic Colloidal Particles, Phys. Rev. Lett. 94(2005) 018301. https://doi.org/10.1103/PhysRevLett.94.018301

[29] Giuseppe Soligno, Marjolein Dijkstra, and Rene van Roij. Self-Assembly of Cubes into 2D Hexagonal and Honeycomb Lattices by Hexapolar Capillary Interactions, Phys. Rev. Lett. 116(2016) 258001. https://doi.org/10.1103/PhysRevLett.116.258001

[30] Yoon Lee, Self-Assembly and Nanotechnology: A Force Balance Approach, 1st edition, Wiley, 2007

[31] Dimitris Stamou, Claus Duschl, and Diethelm Johannsmann, Long-range attraction between colloidal spheres at the air-water interface: The consequence of an irregular meniscus, Phys. Rev. E 62(2000) 5263. https://doi.org/10.1103/PhysRevE.62.5263

[32] W.A. Gifford and L.E. Scriven, On the attraction of floating particles, Chemical Engineering Science 26(1971) 287-297. https://doi.org/10.1016/0009-2509(71)83003-8

[33] Shemaiah M. Weekes, Feodor Y. Ogrin, William A. Murray and Paul S. Keatley, Macroscopic Arrays of Magnetic Nanostructures from Self-Assembled Nanosphere Templates, Langmuir 23(2007) 1057-1060. https://doi.org/10.1021/la061396g

[34] P.A Kralchevsky, V.N Paunov, I.B Ivanov and K Nagayama, Capillary meniscus interaction between colloidal particles attached to a liquid—fluid interface, J. Colloid Interface Sci., 151(1992) 79-94. https://doi.org/10.1016/0021-9797(92)90239-I

[35] V.N. Paunov, P.A. Kralchevsky, N.D. Denkov, I.B. Ivanov and K. Nagayama, Capillary meniscus interaction between a microparticle and a wall, Colloids Surf., 67(1992), 119-138. https://doi.org/10.1016/0166-6622(92)80292-A

[36] P. A. Kralchevsky, V. N. Paunov, and Kuniaki Nagayama, Lateral capillary interaction between particles protruding from a spherical liquid layer, J. Fluid Mech. 299(1995) 105-132. https://doi.org/10.1017/S0022112095003442

[37] Krassimir D Danov, Peter A Kralchevsky, Boris N Naydenov and Günter Brenn, Interactions between particles with an undulated contact line at a fluid interface: capillary multipoles of arbitrary order, J. Colloid Interface Sci., 287(2005), 121-134. https://doi.org/10.1016/j.jcis.2005.01.079

[38] Iris B. Liu, Nima Sharifi-Mood and Kathleen J. Stebe, Capillary Assembly of Colloids: Interactions on Planar and Curved Interfaces, Annu. Rev. Condens. Matter Phys., 9(2018), 283-305. https://doi.org/10.1146/annurev-conmatphys-031016-025514

[39] Jean-Christophe Loudet and Bernard Pouligny, How do mosquito eggs self-assemble on the water surface?. Eur. Phys. J. E 34(2011). https://doi.org/10.1140/epje/i2011-11076-9.

[40] Peter A. Kralchevsky, Nikolai D. Denkov, Krassimir D. Danov, Particles with an Undulated Contact Line at a Fluid Interface: Interaction between Capillary Quadrupoles and Rheology of Particulate Monolayers, Langmuir 17(2001), 7694-7705, https://doi.org/10.1021/la0109359

[41] J.B. Fournier and P. Galatola, Anisotropic capillary interactions and jamming of colloidal particles trapped at a liquid-fluid interface, Phys. Rev. E 65(2002), 031601. https://doi.org/10.1103/PhysRevE.65.031601

[42] G. Morris, S.J. Neethling, and J.J. Cilliers, An investigation of the stable orientations of orthorhombic particles in a thin film and their effect on its critical failure pressure, J. Colloid Interface Sci., 361(2011) 370-380. https://doi.org/10.1016/j.jcis.2011.05.061

[43] Jinbo He, Qingling Zhang, Suresh Gupta, Todd Emrick, Thomas P. Russell and Pappannan Thiyagarajan, Drying Droplets: A Window into the Behavior of Nanorods at Interfaces, Small 3(2007) 1214-1217 https://doi.org/10.1002/smll.200700055

[44] Jacob Lucassen, Capillary forces between solid particles in fluid interfaces, Colloids Surf. 65(1992) 131-137. https://doi.org/10.1016/0166-6622(92)80268-7

[45] Lu Yao, Lorenzo Botto, Marcello Cavallaro, Jr, Blake J. Bleier, Valeria Garbin and Kathleen J. Stebe, Near field capillary repulsion, Soft Matter, 9(2013), 779-786. https://doi.org/10.1039/C2SM27020J

[46] Ping Liu, Julius W. J. de Folter, Andrei V. Petukhov and Albert P. Philipse, Reconfigurable assembly of superparamagnetic colloids confined in thermo-reversible microtubes, Soft Matter, 2015, 11, 6201-6211. https://doi.org/10.1039/C5SM01035G

[47] David G Grier, Optical tweezers in colloid and interface science, Curr. Opin. Colloid Interface Sci., 2(1997) 264-270. https://doi.org/10.1016/S1359-0294(97)80034-9

[48] Bartosz A. Grzybowski, Krzysztof Fitzner, Jan Paczesny and Steve Granick, From dynamic self-assembly to networked chemical systems, Chem. Soc. Rev. 46(2017), 5647-5678. https://doi.org/10.1039/C7CS00089H

[49] Markus Niederberger, Multiscale Nanoparticle Assembly: From Particulate Precise Manufacturing to Colloidal Processing, Adv. Funct. Mater. 2017, 27, 1703647. https://doi.org/10.1002/adfm.201703647

[50] Marek Grzelczak, Jan Vermant, Eric M. Furst, Luis M. Liz-Marzán, Directed Self-Assembly of Nanoparticles, ACS Nano. 4(2010) 3591- 3605. https://doi.org/10.1021/nn100869j

[51] K. Zahn, R. Lenke, and G. Maret, Two-Stage Melting of Paramagnetic Colloidal Crystals in Two Dimensions, Phys. Rev. Lett. 82(1999) 2721. https://doi.org/10.1103/PhysRevLett.82.2721

[52] Fernando Martínez-Pedrero, Francisco Ortega, Joan Codina, Carles Calero and Ramón G Rubio, Controlled disassembly of colloidal aggregates confined at fluid interfaces using magnetic dipolar interactions, J. Colloid Interface Sci., 560(2020) Journal of Colloid and Interface Science 388-397. https://doi.org/10.1016/j.jcis.2019.10.008

[53] E C Mbamala and H H von Grünberg, Effective interaction of a charged colloidal particle with an air-water interface, J. Phys.: Condens. Matter 14(2002) 4881. https://dx.doi.org/10.1088/0953-8984/14/19/313

[54] Shang Hao Piao, Seung Hyuk Kwon, Wen Ling Zhang and Hyoung Jin Choi, Celebrating Soft Matter's 10th Anniversary: Stimuli-responsive Pickering emulsion polymerized smart fluids, Soft Matter 11(2015) 646-654. **https://doi.org/10.1039/C4SM02393E**

[55] Fernando Bresme and Nicholas Quirke, Computer simulation of wetting and drying of spherical particulates at a liquid–vapor interface, J. Chem. Phys. 110(1999), 3536–3547. https://doi.org/10.1063/1.478221

[56] John David Jackson, Classical Electrodynamics, 3rd edition, Wiley, New York, 1999.

[57] Fernando Martinez-Pedrero, Andrejs Cebers, and Pietro Tierno, Dipolar Rings of Microscopic Ellipsoids: Magnetic Manipulation and Cell Entrapment, Phys. Rev. Appl., 6 (2016) 034002. https://doi.org/10.1103/PhysRevApplied.6.034002.

[58] Jianping Ge, Le He,Yongxing Hu and Yadong Yin, Magnetically induced colloidal assembly into field-responsive photonic structures, Nanoscale, 3 (2011) 177-83. https://doi.org/10.1039/C0NR00487A

[59] S. Sacanna and A. P. Philipse, Preparation and Properties of Monodisperse Latex Spheres with Controlled Magnetic Moment for Field-Induced Colloidal Crystallization and (Dipolar) Chain Formation, Langmuir, 22(2006) 10209–10216. https://doi.org/10.1021/la0616505

[60] Bethany J. Newton, Kenneth A. Brakke, and D. Martin A. Buzza. Phys. Chem. Chem. Phys. 16(2014) 26051-26058. https://doi.org/10.1039/C4CP04270K

[61] S. Cappelli 2016 PhD thesis (Eindhoven University, Department of Biomedical Engineering).

[62] Stefano Cappelli, Arthur M. de Jong, Jean Baudry and Menno W. J. Prins, Interfacial rheometry of polymer at a water–oil interface by intra-pair magnetophoresis, Soft Matter, 12(2016) 5551-5562. http://dx.doi.org/10.1039/C5SM02917A

[63] Alexander van Reenen, Yang Gao, Arjen H. Bos, Arthur M. de Jong, Martien A. Hulsen, Jaap M. J. den Toonder and Menno W. J. Prins, Accurate quantification of magnetic particle properties by intra-pair magnetophoresis for nanobiotechnology, Appl. Phys. Lett. 103(2013) 043704. https://doi.org/10.1063/1.4816429

[64] Scott S. H. Tsai, Jason S. Wexler, Jiandi Wan and Howard A. Stone, Microfluidic ultralow interfacial tensiometry with magnetic particles, Lab Chip, 13(2013), 119-125. https://doi.org/10.1039/C2LC40797C

[65] Weijia Wen, Lingyun Zhang, and Ping Sheng, Planar Magnetic Colloidal Crystals, Phys. Rev. Lett. 85(2000), 5464-5467. https://doi.org/10.1103/PhysRevLett.85.5464

[66] G. Grosjean, M. Hubert, and N. Vandewalle, Magnetocapillary self-assemblies: Locomotion and micromanipulation along a liquid interface, Adv. Colloid Interface Sci., 255(2018), 84-93. https://doi.org/10.1016/j.cis.2017.07.019

[67] S. Lefebure, C. Ménager, V. Cabuil, M. Assenheimer, F. Gallet and C. Flament, Langmuir Monolayers of Monodispersed Magnetic Nanoparticles Coated with a Surfactant, J. Phys. Chem. B, 102(1998), 2733- 2738. https://doi.org/10.1021/jp980403+

[68] M. Golosovsky, Y. Saado, and D. Davidov, Energy and symmetry of self-assembled two-dimensional dipole clusters in magnetic confinement, Phys. Rev. E, 65(2002), 061405. https://link.aps.org/doi/10.1103/PhysRevE.65.061405

[69] Koohee Han, Gašper Kokot, Shibananda Das, Roland G. Winkler, Gerhard Gompper and Alexey Snezhko, Reconfigurable structure and tunable transport in synchronized active spinner materials, Sci. Adv. 6(2020), eaaz8535. https://doi.org/10.1126/sciadv.aaz8535

[70] Guido Avvisati, Teun Vissers, and Marjolein Dijkstra, Self-assembly of patchy colloidal dumbbells, J. Chem. Phys. 142(2014), 084905. https://doi.org/10.1063/1.4913369

[71] A. Snezhko, I. S. Aranson, and W.-K. Kwok, Dynamic self-assembly of magnetic particles on the fluid interface: Surface-wave-mediated effective magnetic exchange, Phys. Rev. E 73(2006), 041306. https://doi.org/10.1103/PhysRevE.73.041306

[72] Bethany J. Newton and D. Martin A. Buzza, Magnetic cylindrical colloids at liquid interfaces exhibit non-volatile switching of their orientation in an external field, Soft Matter 12(2016), 5285-5296. https://doi.org/10.1039/C6SM00136J

[73] F. Martínez-Pedrero, A. González-Banciella, A. Camino, A. Mateos-Maroto, F. Ortega, R. G. Rubio, I. Pagonabarraga and C. Calero, Static and Dynamic Self-Assembly of Pearl-Like-Chains of Magnetic Colloids Confined at Fluid Interfaces, Small 17(2021), 2101188. https://doi.org/10.1002/smll.202101188

[74] Gary B. Davies, Timm Krüger, Peter V. Coveney, Jens Harting and Fernando Bresme, Interface deformations affect the orientation transition of magnetic ellipsoidal particles adsorbed at fluid–fluid interfaces, Soft Matter 10(2014), 6742-6748. https://doi.org/10.1039/C4SM01124D

[75] G.B. Davies, T Krüger, P.V. Coveney, J Harting and F. Bresme, Assembling Ellipsoidal Particles at Fluid Interfaces Using Switchable Dipolar Capillary Interactions. Adv. Mater., 26 (2014), 6715-6719. https://doi.org/10.1002/adma.201402419.

[76] SH. Kim, S. Lee, SM. Yang and Gi-Ra Yi. Self-assembled colloidal structures for photonics. NPG Asia Mater 3(2011), 25–33. https://doi.org/10.1038/asiamat.2010.192.

[77] Daniel B. Allan, Daniel M. Firester, Victor P. Allard, Daniel H. Reich, Kathleen J. Stebe and Robert L. Leheny, Linear and nonlinear microrheology of lysozyme layers forming at the air–water interface. Soft Matter, 10(2014) , 7051-7060. https://doi.org/10.1039/C4SM00484A

[78] Prajnaparamita Dhar, Yanyan Cao, Thomas M. Fischer, and J. A. Zasadzinski, Active Interfacial Shear Microrheology of Aging Protein Films, Phys. Rev. Lett. 104 (2010) 016001. https://doi.org/10.1103/PhysRevLett.104.016001

[79] Wendong Wang, Joshua Giltinan, Svetlana Zakharchenko and Metin Sitti, Dynamic and programmable self-assembly of micro-rafts at the air-water interface. Sci. Adv. 3(2017) , e1602522. https://doi.org/10.1126/sciadv.1602522

[80] Alexander Mauro, Dielectrophoresis: The Behavior of Neutral Matter in Nonuniform Electric Fields. Herbert A. Pohl. Q. Rev. Biol. 55(1980), 68-69. https://doi.org/10.1086/411635

[81] Alice P. Gast and Charles F. Zukoski, Electrorheological fluids as colloidal suspensions, Adv. Colloid Interface Sci. 30(1989), 153-202. https://doi.org/10.1016/0001-8686(89)80006-5

[82] Richard W. OBrien and Lee R. White, Electrophoretic mobility of a spherical colloidal particle, J. Chem. Soc. Faraday Trans. 2, 74(1978), 1607-1626. https://doi.org/10.1039/F29787401607

[83] Seth Fraden, Alan J. Hurd, and Robert B. Meyer, Electric-field-induced association of colloidal particles, Phys. Rev. Lett. 63(1989) 2373. https://doi.org/10.1103/PhysRevLett.63.2373

[84] Frieder Mugele and Jean-Christophe Baret, Electrowetting: from basics to applications, J. Phys.: Condens. Matter 17(2005) R705. https://dx.doi.org/10.1088/0953-8984/17/28/R01

[85] N. Aubry, P. Singh, M. Janjua, and S. Nudurupati PNAS 105(2008) 3711-3714. https://doi.org/10.1073/pnas.0712392105

[86] Lionel Foret and Alois Würger, Electric-Field Induced Capillary Interaction of Charged Particles at a Polar Interface Phys. Rev. Lett. 92(2004) 058302. https://doi.org/10.1103/PhysRevLett.92.058302

[87] M. G. Nikolaides, A. R. Bausch, M. F. Hsu, A. D. Dinsmore, M. P. Brenner, C. Gay and D. A. Weitz, Electric-Field-Induced Capillary Attraction between Like-Charged Particles at Liquid Interfaces, Nature 420(2002) 299–301. https://doi.org/10.1038/nature01113

[88] Nadine Aubry and Pushpendra Singh. Physics underlying controlled self-assembly of micro- and nanoparticles at a two-fluid interface using an electric field, Phys. Rev. E 77(2008) 056302. https://doi.org/10.1103/PhysRevE.77.056302

[89] N Aubry and P Singh, Control of electrostatic particle-particle interactions in dielectrophoresis, Europhys. Lett. 74(2006) 623. https://dx.doi.org/10.1209/epl/i2005-10569-0

[90] Peter A. Kralchevsky and Nikolai D. Denkov, Capillary forces and structuring in layers of colloid particles, Curr. Opin. Colloid Interface Sci. 6(2001) 383-401. https://doi.org/10.1016/S1359-0294(01)00105-4

[91] Bartosz A. Grzybowski, Krzysztof Fitzner, Jan Paczesny and Steve Granick, From dynamic self-assembly to networked chemical systems, Chem. Soc. Rev. 46(2017) 5647-5678. https://doi.org/10.1039/C7CS00089H

[92] Jérôme J. Crassous, Adriana M. Mihut, Erik Wernersson, Patrick Pfleiderer, Jan Vermant, Per Linse and Peter Schurtenberger, Field-induced assembly of colloidal ellipsoids into well-defined microtubules. Nat Commun 5(2014), 5516. https://doi.org/10.1038/ncomms6516

[93] J. Kadaksham, Praval Singh, and Nadine Aubry, Manipulation of particles using dielectrophoresis, Mech. Res. Commun. 33(2006) 108-122. https://doi.org/10.1016/j.mechrescom.2005.05.017

[94] Sai Nudurupati, Mohammad Janjua, Nadine Aubry and Pushpendra Singh, Concentrating particles on drop surfaces using external electric fields, Electrophoresis, 29(2008)1164-1172. https://doi.org/10.1002/elps.200700676

[95] Kyuho Hwang, Pushpendra Singh, and Nadine Aubry, Destabilization of Pickering emulsions using external electric fields 2010 Electrophoresis 31(2010) 850-859. https://doi.org/10.1002/elps.200900574

[96] Sai Nudurupati, Muhammad Janjua, Pushpendra Singh and Nadine Aubry, Effect of parameters on redistribution and removal of particles from drop surfaces, Soft Matter 6(2010) 1157-1169. https://doi.org/10.1039/B912723B

[97] P. Dommersnes, Z. Rozynek, A. Mikkelsen, R. Castberg, K. Kjerstad, K. Hersvik & Jon Otto Fossum, Active structuring of colloidal armour on liquid drops. Nat Commun 4 (2013)2066. https://doi.org/10.1038/ncomms3066.

[98] Geoffrey Ingram Taylor, A. D. McEwan, and L. N. J. de Jong. 1966 *Proceedings of the Royal Society of London. Series A. Mathematical and Physical Sciences* 291(1966) 159-166.

[99] M. Janjua, S. Nudurupati, I. Fischer, P. Singh, N. Aubry, Electric field induced alignment and self-assembly of rods on fluid–fluid interfaces, Mech. Res. Commun. 36(2009) 55-64.https://doi.org/10.1016/j.mechrescom.2008.07.007

[100] Alena Sergeeva, Dmitry Gorin, and Dmitry Volodkin, Polyelectrolyte Microcapsule Arrays: Preparation and Biomedical Applications, BioNanoSci. 4(2013) 1-14. https://doi.org/10.1007/s12668-013-0121-6

[101] F. Martínez-Pedrero, Static and dynamic behavior of magnetic particles at fluid interfaces, Adv. Colloid Interface Sci., 284 (2020) 102233. https://doi.org/10.1016/j.cis.2020.102233

[102] Pushpendra Singh, Daniel D. Joseph and Nadine Aubry, Dispersion and attraction of particles floating on fluid–liquid surfaces, Soft Matter, 6(2010), 4310-4325. https://doi.org/10.1039/C000495M

[103] Stefano Cappelli, Arthur M. de Jong, Jean Baudry and Menno W. J. Prins, Interparticle Capillary Forces at a Fluid–Fluid Interface with Strong Polymer-Induced Aging, *Langmuir* 33(2017) 696-705. https://doi.org/10.1021/acs.langmuir.6b03910

[104] Ville Liljeström, Chen Chen, Paul Dommersnes, Jon Otto Fossum and André H. Gröschel, Active structuring of colloids through field-driven self-assembly, Curr. Opin. Colloid Interface Sci. 40(2019) 25-41. https://doi.org/10.1016/j.cocis.2018.10.008

[105] Muhammad Janjua, Sai Nudurupati, Pushpendra Singh and Nadine Aubry, Electric field-induced self-assembly of micro- and nanoparticles of various shapes at two-fluid interfaces, Electrophoresis 32(2011) 518-526. https://doi.org/10.1002/elps.201000523

# Figures

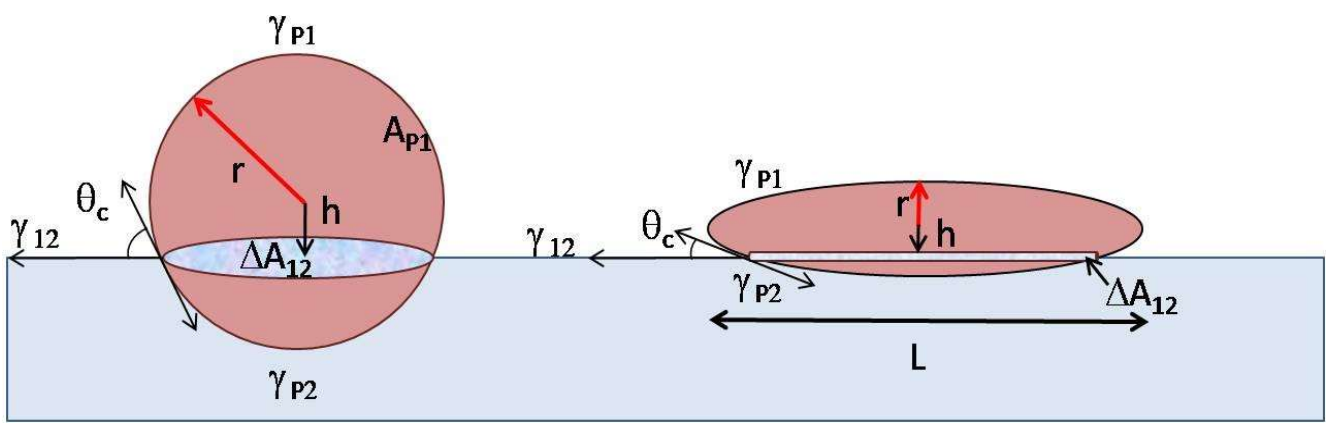

**Figure 1:** A sphere and ellipsoid at fluid-fluid interface. *h* is the height of the particle center of mass above the interface

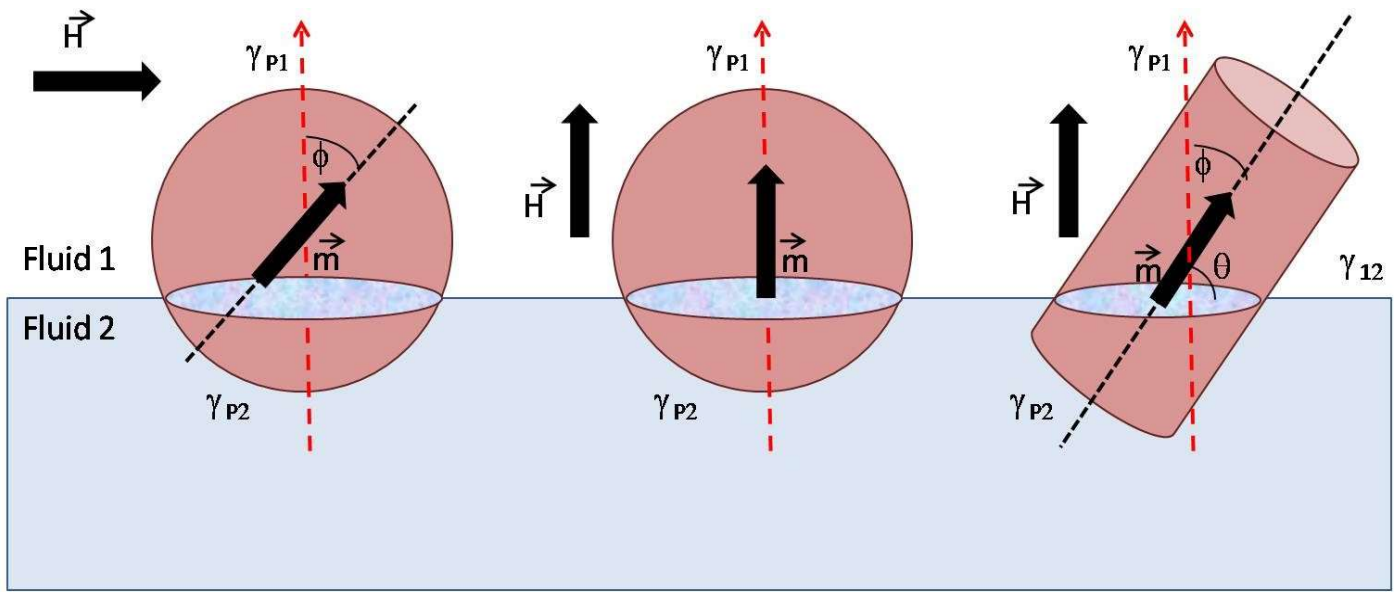

**Figure 2 :** Spherical and cylindrical particles at fluid interfaces showing the relevant parameters

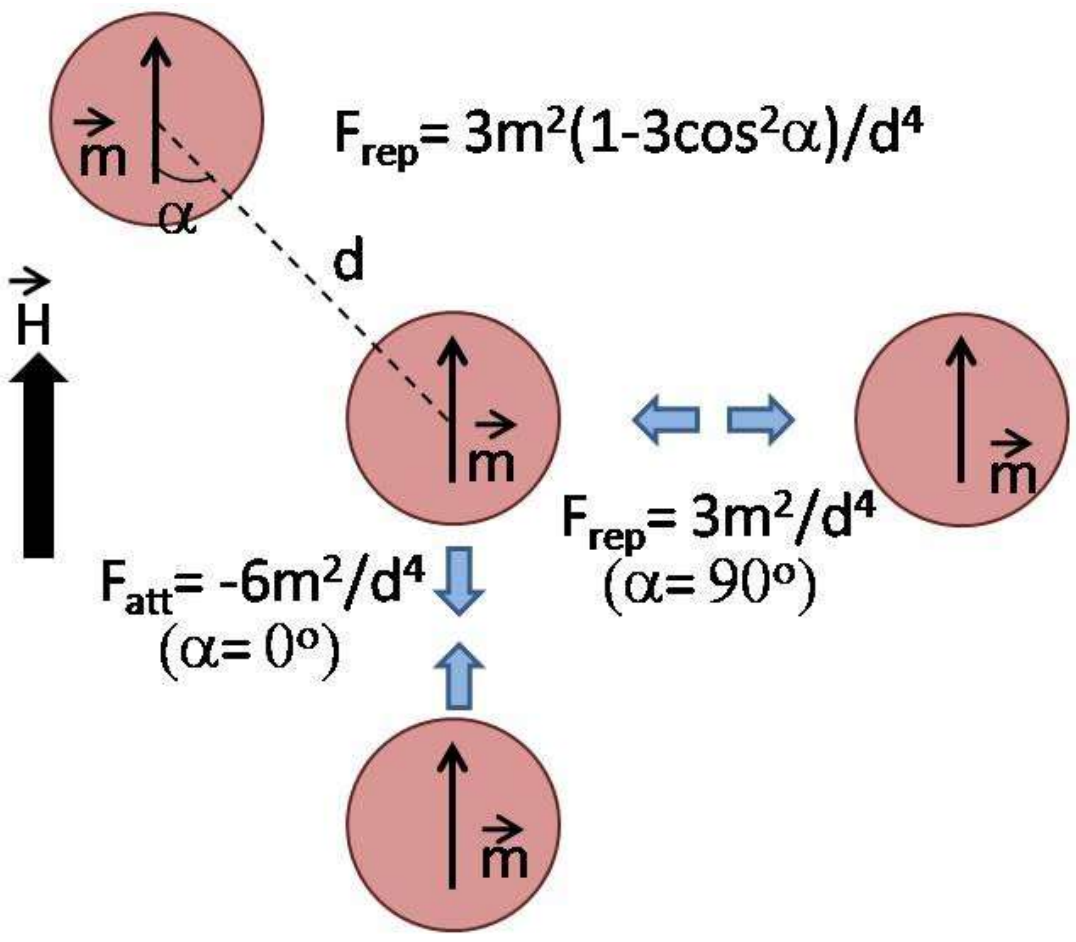

**Figure 3**: Interaction between particles in a magnetic field

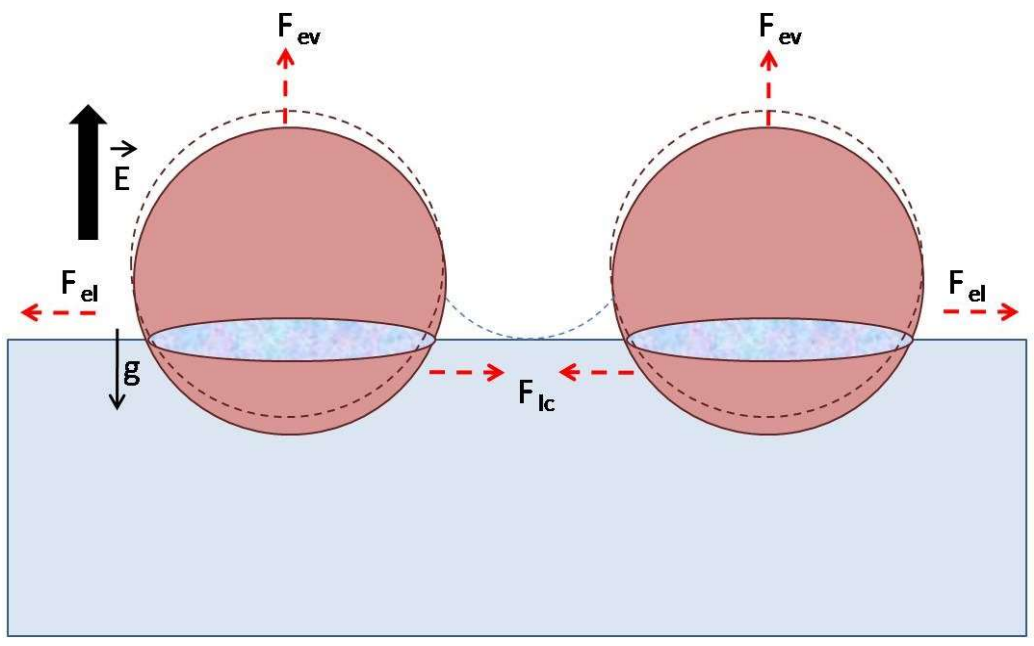

**Figure 4**: Various forces acting on a pair of spheres at interface under perpendicular electric field.